\documentclass[12pt]{article}
\usepackage[T1]{fontenc}
\usepackage{graphics}
\usepackage{eepic}
\usepackage{graphicx}
\input{epsf}

\begin{document}
\centerline {{\bf Monte Carlo simulations of the inside-intron
recombination}}

\vspace{6ex}

\noindent
Stanis{\l}aw Cebrat,\\
 Department of Genomics, Wroc{\l}aw University,
ul. Przybyszewskiego 63/77, 51-148 Wroc{\l}aw, Poland;
cebrat@microb.uni.wroc.pl,\\
Andrzej P\c{e}kalski,\\
 Institute of Theoretical Physics,
Wroc{\l}aw University, pl. Maxa Borna 9, 50-204 Wroc{\l}aw,
Poland;
apekal@ift.uni.wroc.pl\\
 Fabian Scharf,\\
  Institute of Theoretical
Physics, Cologne University, D-50923 Cologne, Germany;
fascharf@gmail.com,

\vspace{5ex}

{\it Abstract}

 Biological genomes are divided into coding and
non-coding regions. Introns are non-coding parts within genes,
while the remaining non-coding parts are intergenic sequences. To
study the evolutionary significance of recombination inside introns
we have used two models based on the Monte Carlo method. In our
computer simulations we have implemented the internal structure of
genes by declaring the probability of recombination between exons.
One situation when inside-intron recombination is advantageous is
recovering functional genes by combining proper exons dispersed in
the genetic pool of the population after a long period without
selection for the function of the gene. Populations have to pass
through the bottleneck, then. These events are rather rare and we
have expected that there should be other phenomena giving profits
from the inside-intron recombination. In fact we have found that
inside-intron recombination is advantageous only in the case when
after recombination, besides the recombinant forms, parental
haplotypes are available and selection is set already on
gametes.\\[5ex]

{\bf Introduction}

 The main forces driving the evolution of living,
sexually reproducing systems, are mutations and recombinations
followed by selection. Mutations are often implemented into the
Monte Carlo models simulating the population evolution as
deleterious events which change the correct or functional gene
(called by geneticists a wild form of the gene) into a
nonfunctional one, just by changing the value of a bit from 0 to
1. In such models genomes are usually represented by bitstrings.
Mutual recombination corresponding to cross-over in biological
system is introduced by choosing randomly the point where pairs of
bitstrings are cut and corresponding fragments replace each other.
In such models recombinations can occur only between genes. This
restriction could produce results significantly different from
natural systems. In fact, in natural genomes genes are not
represented by single bits but by long sequences where
recombination can occur. Furthermore, cross-over is a type of
recombinations which occur in eukaryotic genomes (i.e. not in
bacteria, without a cell nucleus) and these genomes are even more
complicated. In the simplest way the eukaryotic genome could be
described as a large string of nucleotides where only relatively
short regions, called genes, code for proteins. Parts laying
in-between different genes are called intergenic sequences or
intergenic space. Most of genes  consist of short coding parts,
called exons and usually much longer non-coding parts called
introns \cite {gilbert}. The whole information necessary to build
a protein -- the real product of gene -- is included in the exons.
In the human genome, like in many other eukaryotic genomes, exons
make only about 0.02 to 0.03 of the whole genome. For example the
total length of the human gene coding for the clotting factor
VIII, whose defect causes hemophilia, is about 200 000  base pairs
while the total length of several exons building the coding part
of this gene is of the order of 7000 base pairs. On the other
hand, the probability of recombination in a given region varies
being roughly proportional to its length. Thus, building the model
of population evolution considering the recombination events we
should assume the recombination in the intergenic space, inside
introns and inside exons. Since the size of exons is negligible,
we could consider only recombinations in the intergenic space
which reshuffle only the complete genes, and recombination inside
introns which reshuffle additionally exons between recombining
alleles. The reshuffling of exons is considered as a powerful
evolutionary tool increasing the rate of protein evolution
\cite{kolkman}. Considering the structure of interrupted genes it
is important to remember that mutation in only one exon of the
gene is usually deleterious for the whole gene. Nevertheless, in
the models, where genes are represented by single bits, the
mutated genes cannot be repaired by recombination. But introducing
the recombination inside introns, there is a possibility to remove
bad exons and to recover the correct form of genes by
recombination.
\newpage
{\bf Model I}\\
{\it Simulations of the age structured populations -- standard
Penna ageing model}

 In the standard Penna model for biological
ageing, individuals are described by their genomes being strings
of bits of declared length \cite{penna}. Genes in the model are
represented by single bits and have no internal structure. If a
bit is set for 0, it means that it is correct (wild type), while
its value equal to 1 corresponds to the mutated, nonfunctional
version of the gene. In the diploid version, each individual is
represented by two bitstrings. Thus, at the same positions on the
two bitstrings an individual possesses two alleles of the same
gene. The main assumption of the model is that the genes are
switched on chronologically -- in the first time step (''year'')
both alleles in the first locus in the genome are switched on, in
the second time step the second pair of alleles is expressed and
so on. Thus, in the standard version of the model, the number of
genes switched on in the genome corresponds to the age of the
individual. The phenotype of the individual depends on the
declared character of the mutation in the expressed gene. If it is
a recessive mutation, the function of the defective gene can be
complemented by its correct allele -- set for 0, located at the
homologous position on the second string. In such a case, both
alleles at a given position have to be defective to determine a
deleterious phenotype. If the locus is declared a dominant one --
the mutation in any of the two alleles in this locus cannot be
complemented. The effect of the switched-on defective genes on the
individual life span depends on the declared threshold $T$, which
corresponds to the allowable number of expressed deleterious
phenotypic traits. If the number of defective traits already
expressed reaches the $T$ value, the individual dies. If before
dying the individual has reached the reproduction age $R$, it can
produce offspring in each time step with the declared probability
$B$ or it can produce the declared number of offspring. Sexual
reproduction is introduced into the model through mimicking the
production of gametes during meiosis; two bitstrings of the
parental genome exchange homologous fragments at a randomly chosen
position with the probability $C$. One of the two recombined
strings is randomly chosen as a gamete and a mutation is
introduced into a randomly chosen locus with probability $M$. If
at the chosen locus the bit is already set for 1, it stays 1. It
means that there are no reversions, though it is possible to
declare reversion probability. The zygote is formed by joining one
gamete produced by a female with another one produced by a male.
The male individual is randomly chosen. In this way the newborn
comes into being, its sex is established with equal probability
for male or female and the life story repeats itself. There are
only a few parameters:
\begin{enumerate}
\item  $T$ -- the upper limit of expressed phenotypic defects, at
which the individual dies;
 \item $R$ -- minimum reproduction age;
\item$B$ -- birth rate, the number of offspring produced by each
female at reproduction age at each time step;
\item $M$ -- mutation
rate, the number of new mutations introduced into each bitstring
during gamete production;
\item $C$ -- the probability of
cross-over between parental bitstrings during gamete production or
the number of cross-overs.
 \end{enumerate}
 We have chosen the
Penna model for our studies because the results of simulations fit
the age structure of real populations and the structure of the
genetic pool of populations follows the prediction of the
Medewar's hypothesis of ageing -- accumulation of defective genes
expressed during the late periods of life \cite{medewar}.
Additionally, the model enables quantitative estimation of the
parameters describing the quality of populations and their genetic
structure \cite{laszkiewicz}.

 {\it Implementing introns into the
model.
}
 In the standard Penna model, genes have no internal
structure and recombination can occur only in-between the genes.
In our modification we have divided each gene into two exons and
recombination can take place also in-between the two exons of a
single gene. The frequency of such recombination corresponds to
the length of the introns -- the higher inside-intron
recombination rate mimics the longer intron. It was introduced
into the model
through a parameter $p$: \\
 $p$ = 0 means that recombination happens only
in the intergenic sequences (like in the standard model),\\
 p = 1 that
recombination is only inside introns.\\
 Any value of $p$ in the
range $(0,1)$ corresponds to the fraction of recombination events
in the introns in relation to all recombinations. Recombinations
inside the relatively small exons are neglected, thus $1 - p$ is
the probability of recombination between two adjacent genes (in
the intergenic space). Further modifications of the Penna model
are described in detail in the results section.\\[3ex]

\bigskip
{\bf  Model II}\\
 {\it Simulations of populations without age
structure.}

 Simulations of populations in the Penna model lead to
the emerging of the specific gradient of frequency of defective
genes in the genomes. Fractions of defective genes expressed after
the minimum reproduction age are increasing with the increasing
age of an individual. To keep the fraction of mutated genes at the
same level for the whole genomes we have used another model,
without age structure \cite{ap}. In this model, the genetic
structure of individuals also consisted of two bitstrings. Like
in the Model I a pair of two alleles (0 or 1) determines a gene. A
good gene is the one which has one pair (00) on either of the
strings. Mutations and recombinations in the intergenic sequences
and inside introns were introduced like in the previous model, but
the survival probability of an individual $i$ at each time step
was defined by the function:
\begin{equation}
P_i = \exp\left(\frac{-a_i\alpha}{f_i}\right),
\end{equation}
 where $a_i$ is the age of the individual, increased after each Monte Carlo step,
 and $f_i$ is the fitness of the individual $i$  defined
 as the number of proper phenotypic
characters, and $\alpha$ is a parameter determining the selection
pressure.  Here also two parents are needed to produce offspring,
and the procedure of attributing them their genomes is analogous
to Model I. In Model II the quality of the whole genome is
considered at each step, when the surviving probability is
counted. Thus the defective genes (bits set for $1$) are evenly
distributed in the genomes (bitstrings).

\bigskip
{\bf Results and Discussion}\\
 {\it Age distribution and mortality
under different strategy of recombination - Model I}

 Simulations were performed considering four
different strategies of recombination:  1 -- like in the standard
Penna model -- recombination only in the intergenic space; 2 --
recombination only in introns -- that is in-between exons of the
same gene; 3 -- half of recombination events in the intergenic
sequences and half inside introns. In all these versions one
recombination event during the gamete production was introduced.
For comparison, simulations without recombination were also
performed. Populations without recombination were significantly
smaller, while populations with different strategy of
recombinations differed -- the average life span of organisms in
the populations with recombination only in the intergenic space
was the longest, which in this model could be translated as the
higher fraction of population in the reproduction age. Thus, the
recombination in the intergenic space is advantageous when
compared with the recombination inside introns. Accordingly, a
higher mortality, particularly in the later ages, was observed for
populations with recombinations inside introns.

{\it Restoring functional genes}

 Since a gene divided by introns
into several exons is considered as defective if at least one exon
in it is defective, it is obvious that the fraction of defective
genes in the genetic pool is higher than the fraction of defective
exons. One can imagine that a gene which is released from
selection pressure, which means that its function is not important
for surviving in given environmental conditions, can freely
accumulate mutations. After a period long enough, all genes in
such a locus in the genetic pool of the population could be
defective. If such genes are composed of exons, it is obvious that
all genes would be inactivated earlier than all exons. We have
performed the simulations where the first three genes in the
genome were released from selection pressure and after about 2000
MC steps almost all these genes in the population were defective
while about 10\% of exons in the genetic pool were still not
mutated (Fig. 1).

\begin{figure}
\begin{center}
\includegraphics[scale=0.5]{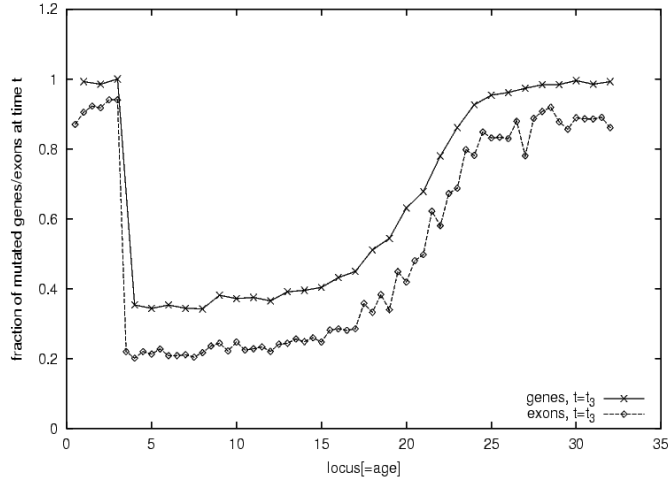}
 \caption{Fractions of mutated exons and of mutated genes after simulations
 without selection pressure on the function of first three genes.
 Parameters of simulation: $R$ = 8, $B$ = 1, $M$ = 1, $C$ =1, $p$ = 0.
 See Model I for explanations of the parameters.} \label{figure1}
\end{center}
\end{figure}

Thus, when selection pressure was set again, it was possible to
restore correct genes by inside-intron recombination. If
recombination in such populations is restricted to the intergenic
sequences, without any possibility of reshuffling the exons --
populations die out. In Fig. 2, the survival probability of such
populations is shown for different $p$ values. At the beginning of
simulations three loci (the same in each genome) were released
from selection pressure and then, after about 100 generations,
selection for the functions of genes in these loci was set again.
As shown in Fig. 2, the probability of surviving of populations
grows with the probability of the recombinations inside introns.
There is still some risk that population would not survive in such
conditions -- it size diminishes substantially after restoring the
selection for the gene function. Such phenomenon is called the
bottleneck effect in biology -- a very risky situation for
population or species. Thus, it is very improbable that all
eukaryotic genes with introns evolved by passing through the
bottleneck. That is why we have looked for other conditions of
evolution when inside-intron recombination could be advantageous.
Since all genes in our simulations are assumed to be composed of
two exons, it is enough to mutate one of the two exons to
eliminate the function of the gene. But if different exons in the
same locus in one diploid genome are defective -- recombination
inside an intron could restore the active gene. Figure 3 describes
the situation; boxes correspond to exons and lines in-between them
to introns where recombination can happen.
\begin{figure}
\begin{center}
\includegraphics[scale=0.5]{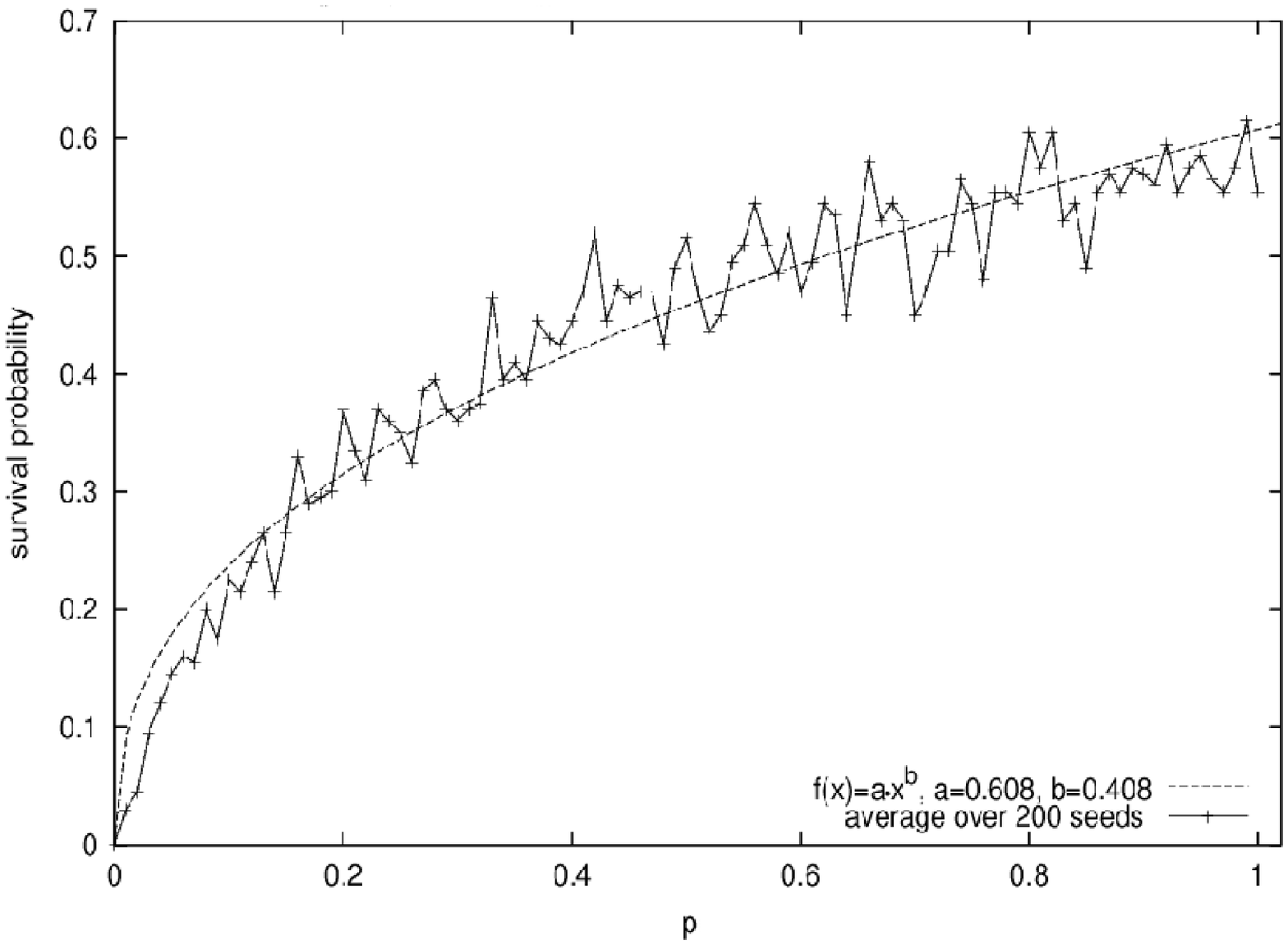}
 \caption{Probability of survival of populations simulated under
 conditions described in Figure 1, after reestablishment of selection
 pressure on the first three gene functions, depending on the
 frequency of recombinations inside introns. Results are averaged over 200
 simulations.} \label{figure2}
\end{center}
\end{figure}
\begin{figure}
\begin{center}
\includegraphics[scale=0.8]{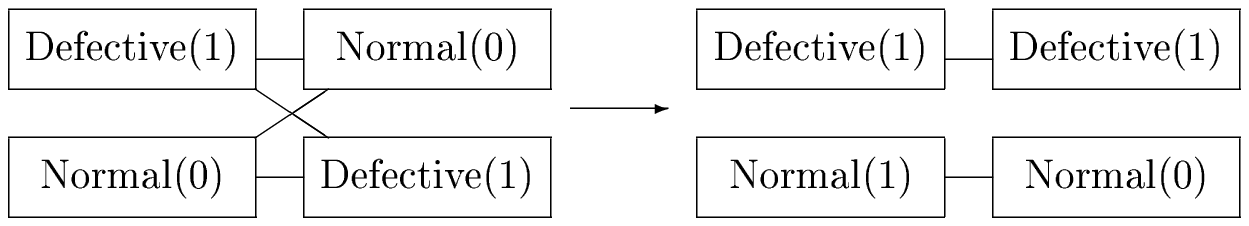}
\caption{Recombination scheme.}
\end{center}
\end{figure}
Unfortunately, recombination could also damage both alleles, if
both exons are deleterious in one allele while in the second
allele both are correct; one only has to reverse the above arrow
to see the situation. It seems that the highest promotion of the
recombination inside introns could be introduced into the model,
if the preselection of gametes is assumed. In the most drastic version
we have assumed that if recombination generates a proper version
of the gene from two defective genes ($01 \times 10 \Rightarrow
00$ and $11$) the gamete $00$ with this correct version of the
gene is selected to form a zygote. Even in such deterministic
conditions, the inside-intron recombination was found to be not
advantageous when compared with intergenic recombination. These
results suggest that the recombination in the second direction
prevails. Looking for the explanation and for conditions when
inside-intron recombination could be profitable, we have noticed
that recombination implemented in the standard Penna model  does
not correspond properly to the meiosis. The standard sexual Penna
model oversimplifies gamete production. In the new modification of
the model, we follow exactly the natural meiosis mechanisms;
before gamete production both haplotypes duplicate giving four
bitstrings and then recombination occurs between pairs of
haplotypes in a randomly chosen point. Since recombination in one
specific site is a rather rare event, it is very improbable that
in both pairs it would occur in the same place. Thus, after
recombination, in respect to the recombination site, among the
four gametes two are recombined and two are in the parental forms.
After introducing this modification of the Penna model and
assuming the preselection of gametes -- recombinations inside
introns are advantageous (see Fig. 4.).

\begin{figure}
\begin{center}
\includegraphics[scale=0.5]{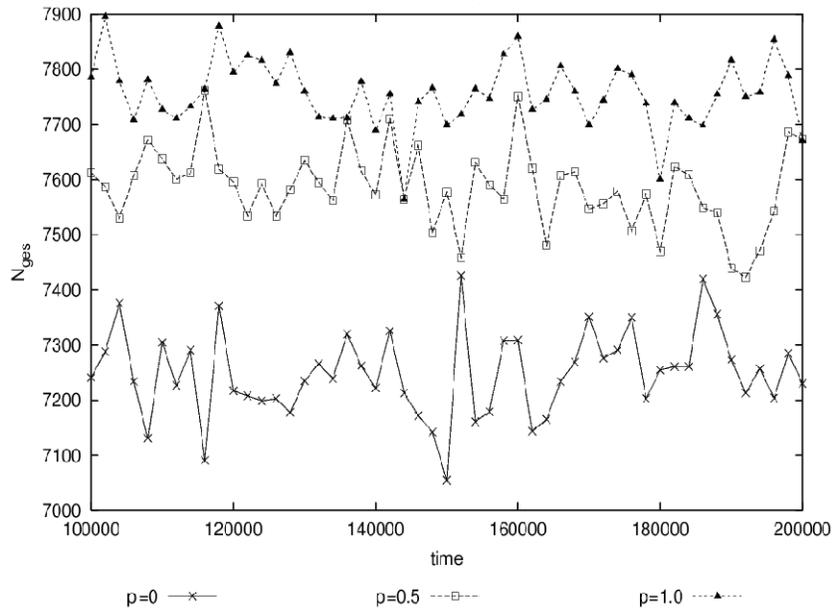}
 \caption{Size of populations with different strategies of
 recombinations. Before gametes production in each of parental
  individuals, two bitstrings replicated and then recombination
 occurred in-between one pair of bitstrings, leaving the other pair
 un-recombined. If recombination inside introns restored the
 correct gene, this gamete was chosen for reproduction.} \label{figure3}
\end{center}
\end{figure}

 In fact, gamete
preselection is very common in Nature. It could be a direct
mechanism of selection like a competition in alternate
haploid/diploid generations like in yeasts or plants, or
haploid/diploid structure of different sex like in the case of
honey bees. Preselection also gave a better justification for
sexual over asexual reproduction \cite{scharf}, \cite{stauffer}.
These results explain also why in some bacteriophages introns
exist while there are no introns in prokaryotic genomes (e.g.
bacterial). Recombinations in the prokaryotic genomes during the
parasexual processes do not leave the parental forms.
Recombinations between phage genomes usually occur when many
genomes are present in one bacterial cell, thus, leaving the
parental forms.

{\it Recombinations inside introns in the populations without age
structure -- Model II.}

\begin{figure}
\begin{center}
\includegraphics[scale=0.5,angle=270]{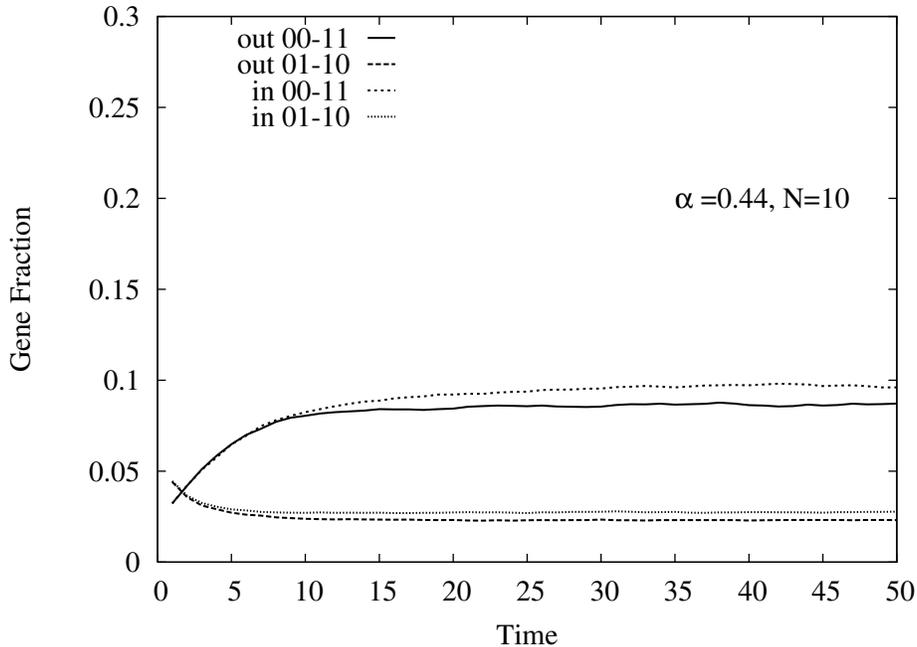}
\caption{Changes in the distribution of defective exons during the
simulation of population evolution.} \label{figure5}
\end{center}
\end{figure}
 To check the role of recombinations inside
introns in the genomes where all genes are switched on in one step
we have used  Model II described in the Models section. In this
model the defective genes are evenly distributed along the genome
and their fraction depends on the mutational and selection
pressures. Using this model we have found, that for shorter
genomes recombination in the intergenic space is a more
advantageous strategy than recombinations inside introns. For
genomes longer than 50 bits, both strategies seem to be equivalent
if the gamete preselection, like in the last version of Penna
model, has been introduced. Since preselection gives the handicap
to the inside-intron recombinations, which could recover the
correct gene from two effective forms, we have checked the
distribution of defective exons in the genetic pool of population
in the equilibrium. To do this we have simulated the populations
until they reach the equilibrium, counted the fractions of
defective exons and then produced the population with random
distribution of the same fraction of defective exons. This
population was let to evolve until it reached the equilibrium.
Notice that the fraction of defective exons has not changed during
this simulation, while their distribution changed (see Fig. 5.).
During the simulation, the fraction of genes with both defective
exons has increased which leads to a higher probability of
recombinations producing both alleles defective. Probability of
recovering the correct alleles from two defective ones is
relatively low - direction of the process to the left prevails in
Figure 3.

\bigskip
{\bf Conclusions}

 We have found that only in very restrictive conditions, with very high gamete
 preselection, the inside-intron recombinations could be advantageous. Maybe,
 under some other combination of parameters, simulations could show conditions
 when such recombination is a better strategy. Nevertheless, in our simulations
 genes could be only in two states: correct or defective. Assuming that positive
 selection can drive the evolution of genes and gene products into the direction
 of higher robustness or more efficient functions, the recombinations inside
 introns could be advantageous as shown by Williams et al., (associated paper
 \cite{williams}).

\end{document}